\documentclass[final]{ws-procs975x65copy}

\begin{document}

\title{Surprises in the phase diagram of\\ the Anderson model on the Bethe lattice}

\author{S. WARZEL \footnote{Plenary lecture given at the International Congress on Mathematical Physics, Aalborg, August 6-11, 2012.}}

\address{Zentrum Mathematik, TU M\"unchen,\\
85747 Garching, Germany\\
E-mail: warzel@ma.tum.de}

\begin{abstract}
The Anderson model on the Bethe lattice is historically among the first for which an energy regime of extended states and a separate regime of localized states could be established. In this paper, we review recently discovered surprises in the phase diagram. Among them is that even at weak disorder, the regime of diffusive transport extends well beyond energies of the unperturbed model into the Lifshitz tails. As will be explained, the mechanism for the appearance of extended states in this non-perturbative regime are disorder-induced resonances.
We also present remaining questions concerning the structure of the eigenfunctions and the associated 
spectral statistics problem on the Bethe lattice. 
\end{abstract}


\bodymatter

\section{A primer to the Anderson model}\label{aba:sec1}

Motivated by the quest for a theory of quantum transport in disordered media, in 1958 Anderson~\cite{A} came up with a model for a quantum particle in a random energy landscape. In its simplest form, the configuration space of the particle is chosen to be $\mathbb{Z}^d $ and the energy landscape is modeled by independent identically distributed (i.i.d.) random variables $  \{ \omega(x) \, | \, x \in \mathbb{Z}^d \} $.  The particles' kinetic and potential energy is described by the operator 
$
H_\lambda(\omega) = T \, + \, \lambda \, V(\omega) 
$
acting in the Hilbert space $ \ell^2(\mathbb{Z}^d) $, where
\begin{itemize}
\item  the first part is the Laplacian without its diagonal terms, i.e.\ $ (T \psi)(x) := -\sum_{|y-x| =1} \psi(y) $, 
\item the second part stands for the multiplication operator corresponding to the random variables, i.e. $ (V(\omega) \psi)(x) := \omega(x) \, \psi(x) $, and
\item the parameter $ \lambda \geq 0 $ was introduced in order to tune the strength of the disorder. 
\end{itemize}
Among the interesting features of this operator is a conjectured dimension-dependent, energetically sharp transition from a regime of localized states to one of diffusive transport, cf.\ Figure~\ref{fig1} below. In order to describe the details of this transition, let us first briefly recall some mathematical framework.\\
\begin{figure}[t]
\begin{center}
\begin{picture}(0,0)%
\includegraphics[height=4cm]{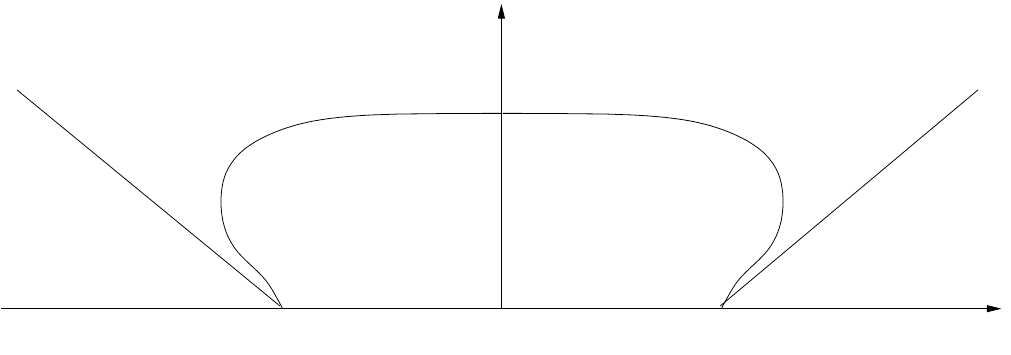}%
\end{picture}%
\setlength{\unitlength}{1993sp}%
%
\begin{picture}(9690,3372)(1189,-3496)
\put(3676,-3436){
{$-2d$}}%
\put(4076,-1900){
{\mbox{\bf ac spectrum}\qquad\qquad{\bf ???}}}%
\put(4076,-2300){
{\mbox{\bf diffusive transport} }}%

\put(7126,-200){
{\mbox{\bf pp spectrum} \qquad $\bf \surd$}}%
\put(7126,-600){
{\mbox{\bf dynamical localization}}}%
\put(11726,-3461){
{$E$}}%
\put(8900,-3436){
{$2d$}}%
\put(5901,-161){
{$\lambda$}}%
\end{picture}%
\end{center}
\caption{Qualitative sketch of the expected phase diagram of the Anderson model for $ d > 2 $ in case $ P_0 $ is a uniform distribution in $[-1,1]$. The outer straight lines mark the edges of the spectrum $ [-2d-\lambda,2d+\lambda]$. The inner curve indicates the conjectured mobility edge separating localized states from delocalized ones for $ d >2 $.}\label{fig1}
\end{figure}
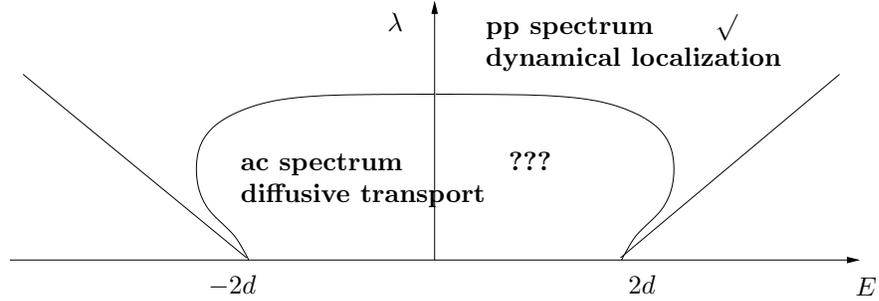

 Since $  H_\lambda(\omega) $ is almost surely (a.s.) self-adjoint, spectral theory is at hand for the analysis of the quantum time evolution $ e^{-itH_\lambda(\omega)} $. 
In this context, it is important to note that the ergodicity of the random process implies that the spectrum of $ H_\lambda(\omega) $, as well as its components in the Lebesgue decomposition are a.s.\ non-random, i.e.\ there is $ \Sigma_\lambda^\# \subset \mathbb{R} $ such that for almost every (a.e.) $ \omega $ one has $ \displaystyle \sigma_\#\big(H_\lambda(\omega)\big) = \Sigma_\lambda^\#   $ with $ \# $ standing for the spectrum, respectively the pure point $ ({\rm pp} )$, singular continuous $(\rm sc) $ and absolutely continuous $({\rm ac}) $ component.  While it is rather straightforward to prove that the spectrum equals $ \Sigma_\lambda = [-2d, 2d] + \lambda\,  {\rm supp}\,  P_0 $, with $ P_0 $ denoting the law of a single variable, determining the spectral components is only trivial in the limiting cases:
\begin{itemize}
\item  the Laplacian has purely ac spectrum in $\sigma(T)= [-2d,2d] $ and its generalized
eigenfunctions are extended plane waves.
\item the random multiplication operator has a.s.\ only (dense) pp spectrum,   $\sigma(V(\omega) ) = \overline{ \{ \omega(x) \, | \, x \in \mathbb{Z}^d \} } =  {\rm supp} \, P_0 $ and its 
	eigenfunctions $\{ \delta_x \, | \, {x\in \mathbb{Z}^d} \}$ are localized at the lattice sites.
\end{itemize}
Within the framework of spectral theory, one may now speak of {\bf spectral localization} respectively {\bf delocalization} within
an energy interval $ I \subset \mathbb{R} $ if the operator $ H_\lambda(\omega) $ has a.s.\ only
 pp spectrum respectively 
only ac (or sc) spectrum there. 
While the RAGE theorem supports some identification of the spectral with dynamical properties, stronger dynamical notions of localization respectively delocalization exist. For example, one speaks of 
strong exponential {\bf  dynamical localization} within $ I \subset \mathbb{R}$  if initially localized wave packets with energies in $ I $ remain localized:
\begin{equation}\label{eq:dynloc}
\sum_{|x|= R} \mathbb{E}\left[\sup_{t\in \mathbb{R}} \big| \langle \delta_x \, , e^{-itH_\lambda} P_I(H_\lambda)\, \delta_0 \rangle \big|^2\right] \leq C \, e^{-R/\xi_I} 
\end{equation}
 where $ \xi_I >0 $ is the energy dependent localization length. Transport on the other hand is often captured by a
{\bf transport index} $ \alpha \in (0,1] $ which describes the asymptotic broadening of  initially localized wave packets with energies in $ I \subset \mathbb{R}$  in the sense that e.g.\ the second moment
		\begin{equation}
		\langle |x(t)|^2\rangle \ :=  \sum_{x\in \mathbb{Z}^d} |x|^2 \;  \mathbb{E}\left[\big| \langle \delta_x \, , e^{-itH_\lambda} P_I(H_\lambda)\, \delta_0 \rangle \big|^2\right] 	
		\end{equation}
grows asymtotically as $ C \, t^{2\alpha} $	 for $ t \to \infty $. Ballistic transport as present in the case $ \lambda = 0 $ corresponds to $ \alpha = 1 $. In case $ \lambda >0 $, one conjectures that 
the transport, if present at all, is always
diffusive, i.e.\ $ \alpha = 1/2 $.

Many other expressions of transport such as a non-vanishing conductivity tensor exist and are investigated.  We refer the interested reader to the textbooks\cite{PF,CL,stoll1} and further articles\cite{ag,Ki} on the subject.\\

Returning to the conjectured spectral and dynamical properties of the operator $ H_\lambda(\omega) $, the situation is summarized as follows:
\begin{itemize}
	\item for $ d = 1, 2 $ one expects spectral (and dynamical) localization throughout the entire spectrum for any $ \lambda > 0 $. 
	\item for $ d > 2 $ the situation is best summarized by the phase diagram in Figure~\ref{fig1}. Its essential feature is a sharp separation of a regime of spectral and strong exponential dynamical localization from that of diffusive transport by a so-called {\bf mobility edge}. 
\end{itemize}
Localization in low dimension ($ d = 1, 2 $) is clearly a non-perturbative quantum phenomenon. While the behavior for $ d = 1 $ has been established by means of transfer matrix methods (starting with the works by Goldsheid, Molchanov and Pastur~\cite{GMP}), establishing complete spectral localization for small $ \lambda >0 $ in dimension $ d = 2$ remains one of the open problems in the field.  

In higher dimensions ($ d > 2 $) localized states occur at the edges of the spectrum, where the states arise from large deviations of the potential, and 
in the region of large disorder in which the features of the multiplication operator dominate. These regimes were made amenable to a mathematical proof through the celebrated multi-scale 
analysis by Fr\"ohlich and Spencer~\cite{FS} who first showed the absence of ac spectrum in these regimes. 
Based on this result, the proof of pp spectrum was later completed by the analysis of Simon and Wolff.\cite{Sim_Wolff} 
A proof of strong exponential localization at band edges and  large disorder was presented by Aizenman in Ref.~\refcite{A_wd}. It uses the fractional moment method developed by Aizenman and Molchanov.\cite{AM} 

Generally one expects that some ac spectrum remains at small disorder in case $ d > 2 $. As already mentioned, these extended states should go along with diffusive transport. Establishing any of these properties remains a huge challenge in the field. 
Steps towards an understanding have been made by Erd\H{o}s, Salmhofer and Yau, 
who proved diffusive behavior in a scaling limit.\cite{ESY} \\

Let us close this short review by briefly mentioning two works related to the quest of an understanding of delocalization of quantum particles in random media. Disertori, Spencer and Zirnbauer have successfully established the existence of a diffusive phase in the supersymmetric hyperbolic non-linear sigma-model.\cite{DSZ} Very recently, Erd\H{o}s, Knowles, Yau and Yin established delocalization and a diffusion profile for the eigenvectors of certain random band matrices.\cite{EKY2}

\section{The phase digram on the Bethe lattice}

\subsection{Some history and the question}
In view of the difficulties of establishing the existence of a regime of delocalization for any high dimensional lattice $ \mathbb{Z}^d $, it is only natural 
to look for simpler models. Historically among the first was the study of the operator $ H_\lambda(\omega) = T + \lambda \, V(\omega) $ in the Hilbert space $ \ell^2(\mathbb{B}) $ over the so-called Bethe lattice. The latter is a regular tree graph, i.e. a graph without loops in which every  
site has the same number $K+ 1 \; (\, \geq 3) $ of neighbors. 
The almost-sure spectrum of the operator  $ H_\lambda(\omega) $ on this graph is easily calculated:
\begin{equation}\label{eq:spectrumB}
\Sigma_\lambda =  [-2\sqrt{K}, 2 \sqrt{K} ]  + \lambda \, {\rm supp} \, P_0 \, .
\end{equation}

It was Abou-Chacra, Anderson and Thouless who first looked at this model and proposed the so-called self-consistency equations (cf.~\eqref{eq:rec} below)  as an approximation to behavior of the self-energy  in high dimensions.\cite{AAT,AT} 
Analyzing these self-consistency equations, they presented convincing arguments for
\begin{itemize}
\item the presistence of a regime of delocalized states for small disorder within the ac spectrum of the Laplacian,  $ \sigma(T) = [-2\sqrt{K}, 2 \sqrt{K} ] $. 
\item the occurrence of localized states for large disorder and outside the interval $[-(K+1),(K+1)] $ for small disorder.  In view of~\eqref{eq:spectrumB}, the latter is a meaningfull statement only if $ {\rm supp} \, P_0 = \mathbb{R} $, i.e. for unbounded random variables.
\end{itemize}
Both of these statements were later proven:  the localization result in the strong sense of~\eqref{eq:dynloc} by Aizenman~\cite{A_wd} and the persistence of ac spectrum alongside ballistic transport by Klein\cite{K,K2}. (Different methods establishing the persistence of ac spectrum can be found in Refs.~\refcite{ASW,FHS}.) 

\begin{figure}[ht]
\begin{center}
\includegraphics[width=.75\textwidth]{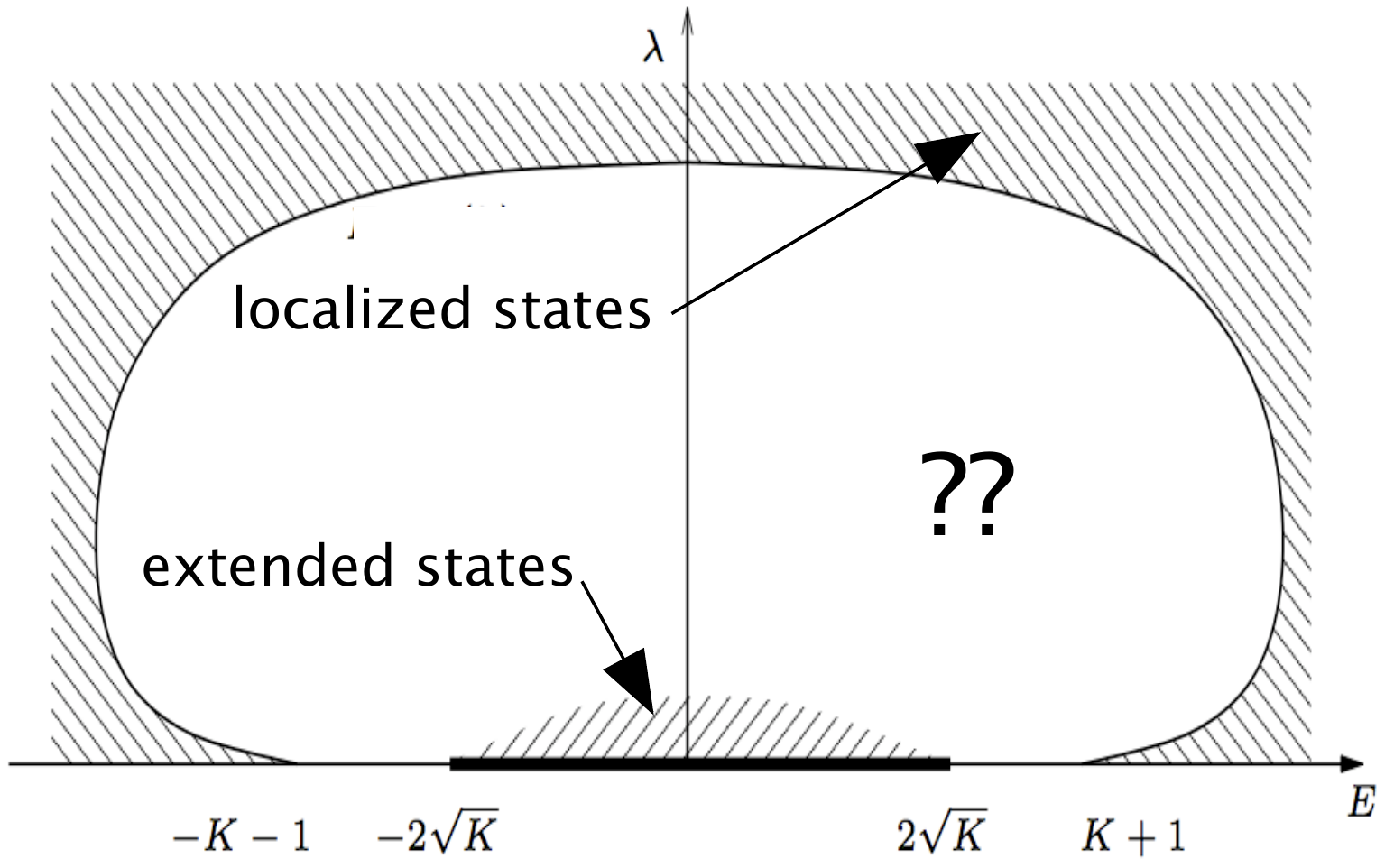}%

\caption{A sketch  of the previously known parts of the phase diagram.  The outer region is of proven localization in the sense of only pp spectrum and strong exponential dynamical localization. The smaller hatched region is of proven  delocalization in the sense of purely ac spectrum and ballistic ($=$ diffusive, cf.\ Subsection~\ref{sub:trans}) transport. The new results address the question of the nature of the states in the middle. }\label{fig2}
\end{center}
\end{figure}

Both of these results left open the question of the location of the mobility gap for small disorder. In particular in case $ {\rm supp} \, P_0 = \mathbb{R} $, the nature of the states in the energy regime $ 2\sqrt{K} < |E | < K+1 $ remained undetermined for small $ \lambda $.  
This question was already taken up by Miller and Derrida~\cite{MD}, but  their (mostly numerical) results remained inconclusive. The situation is summarized in Figure~\ref{fig2}. \\

Triggered by interests in many-particle localization for which the tree is proposed as emulating features of the configuration space of many particles\cite{BAA}, the question of the location of the mobility edge has recently received renewed interest in the physics community. For example, the phase diagram in case $P_0 $ is a uniform distribution on $ [-1,1] $ was numerically investigated in Ref.~\refcite{BST} which presented a diagram resembling the features of the one on the regular lattice (cf.\ Figure~\ref{fig3}).
Before we come to a  critique of the results of Ref.~\refcite{BST} and the solution to the riddle of the location of the mobility edge for small disorder, let us first make a detour into other related results in the literature which present some red herrings.

In case of unbounded random variables, $ {\rm supp} \, P_0 = \mathbb{R} $, the states in the regime of small disorder $ \lambda $ and energies $ |E |> 2 \sqrt{K} $  stem for large and hence rare fluctuations of the random variables. As a consequence, the density of states $ D_\lambda(E) $, i.e. the Lebesgue density of the integrated density of states,\footnote{Note that the existence of a Lebesgue density is guaranteed by a Wegner estimate in case $ P_0 $ has a bounded density.\cite{W}  In fact, if $ P_0 $ is sufficiently close to a Cauchy distribution,  the density of states in known to be analytic.\cite{AK}}
\begin{equation}
	N_\lambda(E) \ := \ \mathbb{E}\left[\langle \delta_0 , P_{(-\infty,E)}(H_\lambda) \, \delta_0 \rangle \right]  = \int_{-\infty}^E \!\!\! D_\lambda(E')\,  dE' \, ,
\end{equation} 
is expected to be severely suppressed. This phenomenon is usually referred to as {\bf Lifshitz tailing}. 
In fact, it is not hard to show that in case of Gaussian random variables, $ N_{\lambda}(E)  \leq \exp{\left(-C(E)/\lambda^2\right)}  $ for any $ E < - 2\sqrt{K} $ and all $ \lambda >0 $.\cite{AW11} Similarly, in case of bounded non-negative random variables $ \omega(x) \geq 0 $, it is conjectured that\cite{BS}
\begin{equation}\label{eq:LA}
\lim_{E\downarrow 0 } \frac{\log \log |\log N_\lambda(-2\sqrt{K} + E) | }{\log E} = - \frac{1}{2} \, . 
\end{equation}
Since the rare fluctuation which cause the suppression of the density of states are typically spatially separated, it is quite natural to suppose that the states in this energy regime are localized in isolated wells. 
As we will point out next, this is not the case and there is a regime of ac spectrum well within the regimes of Lifshitz tails.

\subsection{The location of the mobility edge}

The phase diagram of the Anderson model on the Bethe lattice turns out to be described in terms of 
the {\bf free-energy function}
\begin{equation}
		\varphi_\lambda(s;z) \  :=\  \lim_{|x|\to \infty} \frac{\log \mathbb{E}\left[\left| G_\lambda(0,x;z) \right|^s\right]}{|x|}  \, , 
\end{equation}
which captures the decay properties of the Green function $ G_\lambda(x,y;z) := \langle \delta_x , (H_\lambda - z )^{-1} \delta_y \rangle $.
Based on factorization properties of the Green function, it is proven in Ref.~\refcite{AW11} that this free energy function is well defined, monotone decreasing and convex in $ s \geq 0 $ for all $ z \in \mathbb{C}^+ :=  \{ w \in \mathbb{C} \, |\, {\rm Im}\, w > 0 \} $.  The following properties and quantities are essential for the description of the spectral phase diagram:
\begin{romanlist}
\item For $ s \in (0,1) $ the limits $ |x| \to \infty $ and $ {\rm Im} \, z \downarrow 0 $ exist and commute. 
The boundary limit 
\begin{equation}
 \varphi_\lambda(1;E) = \lim_{s\uparrow 1} \lim_{\eta \downarrow 0} \varphi_\lambda(s;E+i\eta)
 \end{equation}
  exists for Lebesgue almost all $ E \in \mathbb{R} $.
\item The derivative at $ s = 0 $ can be identified with a {\rm Lyapunov exponent} 
\begin{equation}
	L_\lambda(z) \ := \ - \lim_{s\downarrow 0} \frac{\varphi_\lambda(s;z)}{s}  \ = \ - \mathbb{E}\left[ \log |\Gamma_\lambda(0;z)|\right]  \, ,
\end{equation} 
where the last equality results from the factorization of the Green function, and $ \Gamma_\lambda(0;z) :=  \langle \delta_0 , (H^{\mathbb{T}}_\lambda -z)^{-1} \delta_0 \rangle$ denotes the Green function of the operator $ H_\lambda $ restricted to the Hilbert space over a regular rooted tree $ \mathbb{T} $ with root denoted by $ 0 $. 
\end{romanlist}

For the formulation of the result, which was proven in~Ref.~\refcite{AW11}, let us briefly state the assumption on the probability distribution $ P_0 $  which will be assumed for the validity of all of the theorems in this review:
\begin{description}
\item[Assumptions on $P_0$:] The single-site probability measure has a bounded density, i.e. $ P_0(dv) = \varrho(v) dv $ for some $ \varrho \in L^\infty(\mathbb{R}) $, which satisfies:
\begin{romanlist}
\item	it  is bounded relative to its minimal function, 
 i.e.,  for Lebesgue-a.e.\ $ v \in \mathbb{R} $ and some $ c < \infty $:
 \begin{equation}
    \varrho(v) \ \leq \  c \, \inf_{\nu \in (0,1] } (2\nu)^{-1}  \int 1_{ |x-v| \leq \nu }\,  \varrho(x) \, dx \, .
  \end{equation}
\item the moment condition with some $ r > 0 $: 
\begin{equation}\label{eq:mom}
		 \int_\mathbb{R} |v|^r \, \varrho(v) \, dv < \infty \, .
\end{equation}
\end{romanlist}
\end{description}
\vfill

The first theorem determines the location of the mobility edge in terms of $  \varphi_\lambda(1;E)  $. For purely technical reasons this theorem was proven in Ref.~\refcite{AW11} only in the unbounded case. The second part of the theorem holds more generally and was already established in Ref.~\refcite{A_wd}.
\begin{theorem}[Characterization of the mobility edge]\label{thm:phi} 
	If $ \inf_{ |v|\le k } \varrho(v) > 0 $ for all $k<\infty$, the free-energy function
	determines the mobility edge in the sense that 
	\begin{itemize}
	\item for a.e.\ $ E \in \Sigma_\lambda $ if $ \varphi_\lambda(1;E)  \ > \ - \log K  $ then a.s.\
	\begin{equation}\label{eq:acDOS}
	  D^{ac}_{\lambda}(E,\omega)   :=  \frac{1}{\pi}\,  {\rm Im} \,G_\lambda(0,0;E+i0,\omega)  > 0 \, .
	\end{equation}
	\item if $ \varphi_\lambda(1;E)  \ < \ - \log K $ for a.e.\ $ E \in I \subset  \Sigma_\lambda  $, then $ I $ is a regime of strong exponential dynamical localization in the sense of~\eqref{eq:dynloc}. 
	\end{itemize}
\end{theorem}

The left-hand side of \eqref{eq:acDOS} is the density of ac states at energy $ E $. The event $ D_\lambda^{ac}(E,\omega) >0 $ happens with either full probability or probability one. Moreover, by translation invariance, if  $ D_\lambda^{ac}(E,\omega) >0 $ a.s., then the ac density associated to any other sites is stricly positive, $ {\rm Im} \,G_\lambda(x,x;E+i0,\omega)  >0 $ a.s. \\

The above theorem identifies the set $ \{ (E,\lambda) \in \mathbb{R}^2 \, | \, \varphi_\lambda(1;E)  = - \log K \} $ as the location of the mobility edge for the Anderson model on the Bethe lattice and hence paints a fairly complete picture. Unfortunately, due to a lack of regularity results  the question whether this set constitutes a line in the phase diagram remains open. 
In fact, since the function $ \varphi_\lambda(1;E) $ is hard to analyze, we do not employ the above criterion directly to answer the question about the location of the mobility edge for small $ \lambda $.
However, in this regime it is reasonable to expect that the estimate $  \varphi_\lambda(1;E)  \geq - L_\lambda(E) $, which derives from the convexity of the free energy function, is rather tight. The following theorem (which for unbounded random variables may also be viewed as a corollary to Theorem~\ref{thm:phi}, but under stronger hypothesis is proven in Ref.~\refcite{AW11}) summarizes the resulting criterion for extended states. 
\begin{theorem}[Lyapunov exponent criterion]\label{thm:Lyap}
For a.e.\ $E \in \Sigma_\lambda $  the condition $ L_\lambda(E) < \log K  $ implies $ D^{ac}_{\lambda}(E)  > 0 $ a.s.
\end{theorem} 
Based on this theorem, the determination of the mobility edge for small $ \lambda $ now proceeds along the following lines:
\begin{romanlist}
\item In case $ \lambda = 0 $ the Lyapunov exponent can be explicitly calculated (using~\eqref{eq:rec} below). One obtains:
 \begin{equation}\label{eq:L0}
		  L_0(E) \  \left\{ \begin{array}{l@\quad l}
		  	= \log \sqrt{K} \quad & \mbox{if}\;  |E | \leq 2 \sqrt{K} \, , \\  
		 \in \left(\log \sqrt{K} , \log K \right) \quad &\mbox{if}\;   2 \sqrt{K} < |E| < K+1\, ,  \\
		  \geq \log K \quad & \mbox{if}\; |E| \geq K+1 \, . 
		 \end{array} \right.
\end{equation}
As an aside we note that since $ G_0(0,x;z) = G_0(0,0;z) \, e^{-|x| L_0(z)} $,  the explicit expression helps identifying  $ E = \pm (K+1) $ as the threshold for the summability of the unperturbed Green function. In other words, the interval $[-(K+1), K+1] $ is the spectrum of $ T $ on the Banach space $ \ell^1(\mathbb{B}) $.
\item Using the fact that the Lyapunov exponent is the negative real part of a Herglotz function, it is not hard to convince oneself that it is weakly continuous in $ \lambda $ in the sense that for any interval $ I \subset \mathbb{R} $:
\begin{equation}\label{eq:contL}
 \lim_{\lambda \downarrow 0 } \, \int_I L_\lambda(E) \, dE = \int_I L_0(E) \, dE \, . 
 \end{equation}
In view~\eqref{eq:L0} this implies~\cite{AW11} the existence  of  {\bf some ac spectrum within the full interval $  (-(K+1), K+1) $} for small $ \lambda $ and $ {\rm supp} \, \varrho = \mathbb{R} $. \\

In case of Cauchy random variables, i.e. $\varrho(v) = \pi^{-1} (v^2+1)^{-1} $, one can even make a stronger statement since the Lyapunov exponent can be explicitly computed: $ L_\lambda(E) = L_0(E+i\lambda) $. The Lyapunov exponent criterion hence identifies a full interval within $  (-(K+1), K+1) $ as the regime of ac states for small $ \lambda $. Since in this case the ac density is strictly positive for a.e.\ energy in a full interval, by a Simon-Wollf type argument, one may even conclude that the ac spectrum on this interval is pure.\cite{AW11}
\item In case of bounded random variables, e.g.\ $ {\rm supp} \, \varrho = [-1,1] $, we prove~\cite{AW11} the continuity of the Lyapunov exponent at the spectral edge $ E_\lambda := \inf \Sigma_\lambda = -2\sqrt{K} - \lambda$:
\begin{equation}
\limsup_{E \downarrow E_\lambda} L_\lambda(E)  \ \leq \ L_0(E_\lambda - \lambda) \, . 
\end{equation}
As a consequence, for any $0< \lambda<\lambda_{min} := (\sqrt{K}-1)^2/2 $ there is some  $ \delta_\lambda>0 $ such that the Lyapunov condition is satisfied on the full interval $ [E_\lambda , E_\lambda + \delta_\lambda] $ and hence there is 
{\bf purely ac spectrum} occurring at the band edges within the regime of conjectured Lifshitz tails~\eqref{eq:LA}. In particular, this shows that the numerical results in Ref.~\refcite{BST} need to be corrected for small $ \lambda $. 
In view of~\eqref{eq:contL} or the perturbative results of Klein\cite{K}, which cover energies within $ [-2\sqrt{K},2\sqrt{K} ] $, we conjecture that for sufficiently small $ \lambda $ the entire spectrum is purely ac, cf.~Fig.~\ref{fig3}.
\begin{figure}[bht]
\begin{center}
\hspace*{1cm}\includegraphics[width=.85\textwidth]{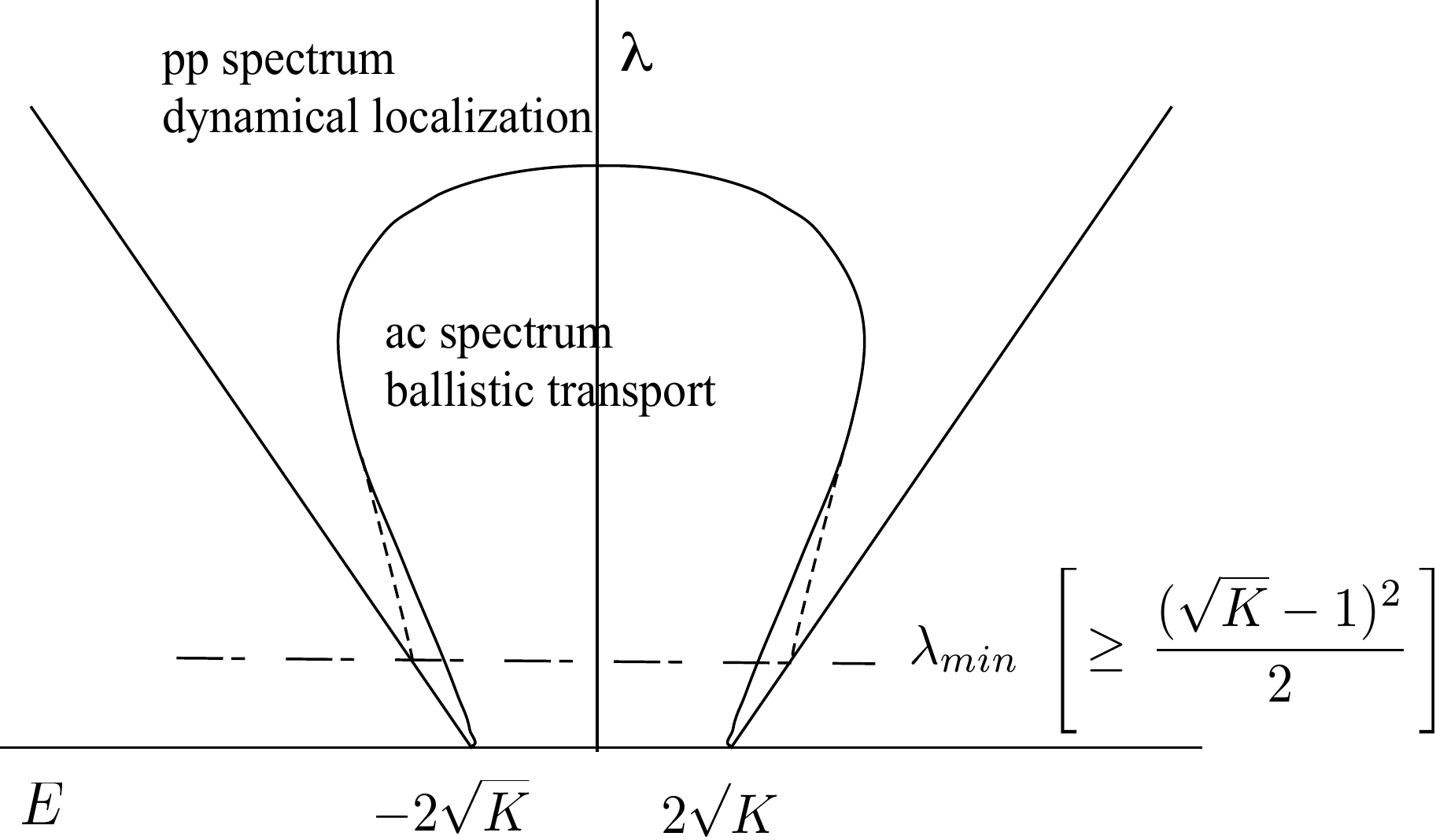}  
\end{center}
\caption{The solid line sketched the previously expected phase diagram for the Anderson model on the Bethe lattice (cf. Ref.\refcite{BST} in case $ P_0 $ is a uniform distribution in $[-1,1] $. Our analysis suggests that at weak disorder there is no localization and the spectrum is purely ac. While the proof of that is incomplete, we do prove that for $\lambda \leq (\sqrt{K} -1)^2/2 $ there is no mobility edge beyond which localization sets in. }\label{fig3}
\end{figure}

\end{romanlist}

 It is of some interest to estimate the {\bf critical disorder strength $ \lambda_{c} > 0 $} above which the entire spectrum is pp and dynamical localization prevails. Using the factorization property of the Green function together with bounds of the form
\begin{equation*}
 \mathbb{E}\left[ |G_\lambda(0,0;z)|^s \, | \, \omega(x) , \, x \neq 0 \right]  \leq   \sup_{a \in \mathbb{C}} \int \frac{\varrho(v)}{|\lambda v - a |^s} \, dv  \leq  \frac{ \|\varrho\|_\infty^s }{(1-s) \, \lambda^s } =: C_s(\lambda) 
 \end{equation*}
 on the conditional fractional moment of the diagonal of  Green functions, conditioned  on all random variables aside from $ \omega(0) $, one obtains 
$\left(  \varphi(1;z) \leq \, \right) \, \varphi(s;z) \leq  \log C_s(\lambda) $. Optimizing over $ s \in(0,1) $, i.e. supposing that $\lambda \geq e \| \varrho \|_\infty $ and picking $ (1 - s)^{-1} = \log( \lambda / \| \varrho\|_\infty )$, we arrive at
\begin{equation}  
\lambda_{c}  \leq  \| \varrho \|_\infty \, K \left( e \, \log K + 1 \right)  \, . 
\end{equation} 
This bound is consistent with the conjectured asymptotics $ \lambda_c \sim C\,  K \log K $ for large~$ K $.

A lower bound on $ \lambda_c $ is only available in case of Cauchy random variables (or, more generally, if the random variables have a Cauchy component). It is based on the Lyapunov exponent criterion and the fact that $ L_\lambda(E) = L_0(E+i\lambda) $. The equality $  L_0(i\lambda) = \log K $ yields $ \lambda_c \geq K-1 $ in this case.

\subsection{Transport properties}\label{sub:trans}

In view of the existence of a regime of extended states within the regime of extremely low density of states, one might wonder whether there is a further dynamical transition if one crosses from the perturbative regime already analyzed by Klein\cite{K2} into that of Lifshitz tails. The following theorem excludes this possibility and establishes ballistic transport throughout the region of positive ac density. 
\begin{theorem}[Ballistic transport in the ac regime]\label{thm:BT}
Assume that~\eqref{eq:mom} holds with $ r = 12 $. Then for any initial state $ \psi = f(H) \delta_0 $, with $ f \in L^2(\mathbb{R}) $ supported in 
$ \left\{  \, E\in \mathbb{R} \, | \,      D^{ac}_{\lambda}(E)  >0   \; \mbox{a.s.} \right\}  $  and all $b>0$:    
\begin{equation}\label{eq:BT}
 \mathbb{E}\Big[  \eta \int_0^\infty \! e^{-\eta t }\! \sum_{|x| <  b/ \eta}  \Big| \big\langle \delta_x \, ,  \, e^{-itH_\lambda} \,  \psi\big\rangle  \Big|^2 \, dt \Big] \ \le \  C(f)  \, b \  +  \ o(\eta)  \, , 
\end{equation}
with some  $C(f) <\infty$, and  $o(\eta)$  a quantity which vanishes for $\eta \to 0$.  
\end{theorem}

Theorem~\ref{thm:BT} in particular implies that within the regime of ac states the time evolution is at least ballistic in the time averaged sense
\begin{equation}
\liminf_{ \eta \downarrow 0 } \eta^3 \int_0^\infty \! e^{-\eta t } \langle |x(t)|^2\rangle \,  dt > 0 \, .
\end{equation}
This proves the ballistic nature of the ac states,
since quite generally quantum evolution is at most ballistic, $ \limsup_{ \eta \downarrow 0 } \eta^3 \int_0^\infty \! e^{-\eta t } \langle |x(t)|^2\rangle \,  dt < \infty $, .  

One might wonder whether this result fits  the general conjecture about the diffusive nature of the transport in random media. For this it is useful to note that on the Bethe lattice the {\bf diffusion is ballistic}:
\begin{equation}\label{eq:diff}
\int_0^\infty \langle \delta_x , \, e^{t \Delta} \delta_0\rangle  \, dt = \langle \delta_x \, , (-\Delta)^{-1} \delta_0 \rangle \ = \ \frac{C}{K^{|x|} }
\end{equation}
where  $ (\Delta \psi)(x) := -T\psi(y) - (K+1) \psi(x) $ stands for the diffusion generator. The proof of Theorem~\ref{thm:BT}, which can be found in Ref.~\refcite{AW12}, essentially proceeds through the following estimate  
on the second moment of Green function
\begin{equation}
 \sup_{\zeta \in I +i(0,1]} \mathbb{E}\left[|G_\lambda(0,x;\zeta)|^2\right] \ \leq \frac{C(I)}{ K^{|x|}}  
\end{equation}
for any $ I \subset  \left\{  \, E\in \mathbb{R} \, | \,      D^{ac}_{\lambda}(E)  >0   \; \mbox{a.s.} \right\}  $. In view of~\eqref{eq:diff} this bound reflects the diffusive behavior.

\section{Resonant delocalization}

We will not attempt to give a proof of any of the results presented in the previous section, but only sketch the ideas behind the geometric resonance mechanism which we dubbed resonant delocalization and which is responsible for the non-perturbative appearance of ac states outside $ \sigma(T) $. 
To simplify the presentation, we restrict ourself to the case of a regular rooted tree $ \mathbb{T} $ with root $ 0 $, for which every site has $ K $ forward neighbors. In this set up, the truncated resolvents
$  \Gamma_\lambda(u;z) :=  \langle \delta_u , (H^{\mathbb{T}_u^+}_\lambda -z)^{-1} \delta_u \rangle $ associated with the subtree $ \mathbb{T}_u^+ $ which is rooted at and forward to $ u $ satisfy the so-called self-consistency equations:
\begin{equation}\label{eq:rec}
\Gamma_\lambda(x;z) = \left( \lambda \omega(x) - z - \sum_{y\succ x} \Gamma_\lambda(y;z) \right)^{-1} \, . 
\end{equation}
Here $H^{\mathbb{T}_u^+}_\lambda $ denotes the restriction of $ H_\lambda $ to $ \ell^2(\mathbb{T}_u^+) $.
The equations~\eqref{eq:rec} consitute a discrete dynamical system system for probability distribution on $ \mathbb{C}^+$ induced by 
the map $ \omega \mapsto \Gamma_\lambda(0;z,\omega)$. 
In order to establish the presence ac spectrum at energy $ E $ within the spectrum, it hence remains to investigate the fixed points of~\eqref{eq:rec} for the boundary values $ z = E+i0 $ and establish the positivity of the right-hand side of 
\begin{equation}\label{eq:recIm}
 {\rm Im}\,  \Gamma_\lambda(0;z)  \geq   \sum_{|x_+ |=R} \left|G_\lambda(0,x;z)\right|^2 \, {\rm Im}\,  \Gamma_\lambda(x_+;z)\, , \qquad  x \prec x_+  \, .
 \end{equation}
Here the sum ranges over the sphere of sites at distance $ R $ from the root in $ \mathbb{T} $ and $ x \prec x_+ $ denotes the unique predecessor of $ x_+ $. \\
We will now present an argument, which proves the following implication:
\begin{itemize}
\item[] If $ {\rm Im} \, \Gamma_\lambda(0,E+i0,\omega) = 0 $ a.s. and $ L_\lambda(E) < \log K $, then there exists some $ \delta > 0 $ such that the event 
\begin{equation}\label{eq:event}
N_R(\omega) := \sum_{|x|=R} 1_{\{ G_\lambda(0,x; E+i0,\omega)  \gtrsim\  e^{\delta  |x|}\} } \ \geq 1 
\end{equation}
occurs with a positive probability, independently how large $ R > 0 $ is.
\end{itemize}
First and foremost, this shows that the Green function cannot be square summable (and hence by a Simon-Wolff argument\cite{Sim_Wolff} there is no pp spectrum). A more detailed analysis based on~\eqref{eq:recIm} establishes the existence of ac spectrum, in contradiction with the assumption in the above implication.\\

 The idea to construct the above enormous fluctuations is based on the following properties of Green functions on trees: 
 \begin{romanlist}
 \item Factorization of the Green function:  $ G_\lambda(0,x; z) \ = \ \widehat G_\lambda(0,x_-; z) \;  G_\lambda(x,x;z) $, where $ \widehat G_\lambda $ pertains to the operator restricted to the tree which is cut beyond $ x $.
\item Local dependence on $ \omega(x) $: \quad $ G_\lambda(x,x;z) \ = \ \left( \lambda \omega(x) - \sigma_\lambda(x;z) \right)^{-1} $, where $ \sigma_\lambda(x;z)  $ is the so-called self-energy  at $ x $ which is independent of $ \omega(x) $. 
\end{romanlist}
 
Since the typical decay of the Green function is given by the Lyapunov exponent, the event  $ {\mathcal{R}_x}:= \{  \widehat G_\lambda(0,x_-;E+i0) \geq\ C  \, e^{-L_\lambda(E) |x|} \} $ occurs asymptotically with probability one.
In order to compensate for this typical decay, we look at the extreme event ${\mathcal{E}_x}:= \{ G_\lambda(x,x;E+i0) \ \geq   e^{(L_\lambda(E)+\delta)  |x|} \} $ which using its local dependence on $\omega(x) $ may be proven to have a conditional probability which is of order $ e^{-(L_\lambda(E)+\delta)  |x|} $. As a consequence, 
$$
\mathbb{E} \left[ N_R \right]  \geq \sum_{|x|=R} \mathbb{P}( \mathcal{R}_x \cap \mathcal{E}_x ) \ \gtrsim C\,  K^R  e^{-(L_\lambda(E)+\delta)  R}  \geq 1 
$$
provided $ L_\lambda(E) < \log K $ and $ \delta >0 $ is sufficiently small.

In order to establish the asymptotic occurrence of the event~\eqref{eq:event}, we use the second moment method which is based on the Paley, Zygmung inequality
\begin{equation}
 \mathbb{P}\left( N_R \geq 1  \right) \ \geq \ \frac{\mathbb{E}\left[N_R\right]^2}{\mathbb{E}\left[N_R^2\right]} \, .
 \end{equation}
An upper bound on the second moment is essentially based on the following consequence of a rank-2 calculation of the conditional expectation, conditioned on all random variables aside from $ x $ and $ y $:
\begin{equation}
		\mathbb{P}\left( \mathcal{E}_x \cap \mathcal{E}_y  \, \big|\,  \omega(\xi) , \xi \not\in \{x, y\} \right) \ \leq \ C \, \tau^{-1} \left(\tau^{-1} + \min\left\{1,  \big| \widehat G(x_-,y_-;E) \big| \right\}\right) \, . 
\end{equation}
Here $ \tau = e^{(L_\lambda(E)+\delta)  |x|} $ and $\widehat G(x_-,y_-;E)  $ denotes the Green function on the subtree of $ \mathbb{T} $  which is chopped beyond $ x $ and $ y $. (For more details, see Ref.~\refcite{AW11}.) 
	
%
%
%
%
%

\section{Some remaining questions}

The above geometric resonance mechanics should also be applicable to homogeneous hyperbolic graphs other than trees. In this context, let us mention that the question of persistence of ac spectra in the perturbative regime has been studied for so-called trees of finite cone-type\cite{KLW}, certain 
decorations of the regular tree\cite{FHH,FHS2,FHS3} 
as well as for the so-called Bethe strip\cite{KSa1,KSa2}. It would be desirable to extend the above analysis to such hyperbolic graphs with loops.\\

Aside from this obvious question about extendability, there is another aspect to the Anderson model on the Bethe lattice which is worth clarifying and which starts to attract attention in the physics community.\cite{DS,BT} This concerns the so-called {\bf spectral statistics} and the associated question of the participation ratio of the eigenfunctions. 

In order to put these questions into come context, let us switch back to the Anderson model on 
$ \mathbb{Z}^d $.  
Carving out a finite cube $ \Lambda \subset \mathbb{Z}^d $, one may study the 
random eigenvalues $ \big\{ E_n(\Lambda,\omega ) \, | \, n \in \mathbb{N} \big\} $ of the restriction of $ H_\lambda(\omega) $ to $ \ell^2(\Lambda) $. The existence of the density of states (which is guaranteed by a Wegner estimate\cite{W}) proves that the average density of these eigenvalues within the spectrum is proportional to the volume $ |\Lambda| $. 
It is therefore reasonable to fix an energy $ E \in \Sigma_\lambda $ and look at the rescaled process of eigenvalues in a window of size $|\Lambda|^{-1} $, i.e. the random measure
\begin{equation}
\mu_{\Lambda}^{E} := \sum_n \delta_{\, |\Lambda | \left(E_n(\Lambda ) - E \right)} \, . 
\end{equation}
The spectral statistics conjecture states that $ \mu_{\Lambda}^{E} $ converges (weakly) as $ \Lambda \uparrow \mathbb{Z}^d $ to
\begin{itemize}
\item a Poisson process if $ E \in \Sigma^{\rm pp}_\lambda $.
\item a GOE process if $ E \in \Sigma^{\rm ac}_\lambda  $. 
\end{itemize}
Here GOE process refers to the (rescaled) process of bulk eigenvalues of a random matrix in the Gaussian Orthogonal Ensemble.
Substituting the assumption $ E \in \Sigma^{\rm pp}_\lambda $ by a fractional-moment localization estimate, the first part of this conjecture was proven by Minami\cite{minami} for arbitrary dimensions and, for $ d = 1 $, earlier by Molchanov\cite{molch}. \\

The question which suggests itself is that of the spectral statistics on the Bethe lattice. Let me first describe how not to approach the spectral-statistics conjecture. 

If one considers a finite regular rooted tree of  $ L  $ generations, i.e. that subtree $\mathbb{T}_L \subset \mathbb{T} $ in which all sites have distance at most $ L \in \mathbb{N} $ from the root, one may look at the eigenvalues $ \{ E_n(L,\omega) \, | \, n \in \mathbb{N} \} $ of the restriction of $ H_\lambda(\omega) $ to $ \ell^2(\mathbb{T}_L) $ and consider the random measure 
\begin{equation}
 \mu_{L}^{E} := \sum_n \delta_{\, |\mathbb{T}_L | \left(E_n(L ) - E \right)} \, . 
 \end{equation} 
It was shown in Ref.~\refcite{AW07} that (under a regularity assumption on the Lyapunov exponent, which e.g. is satisfied for Cauchy random variables) the limiting process is always Poisson, irrespectively whether $ E $ was taken from the regime of localized or delocalized states.
\begin{theorem}[Possion statistics throughout the spectrum] Suppose that the Lyapunov exponent $ L_\lambda(E) $ is an equicontinuous function of $ E \in I $. Then
the rescaled process of eigenvalues $ \mu_{L}^{E} $  converges weakly as $ L\to \infty $ to a Poisson process 
for a.e.\ $ E \in I $. 
\end{theorem}
At first look, this seems surprising. However, as is explained in detail in Ref.~\refcite{AW07}, it does not contradict the spectral-statistics conjecture described above if carefully interpreted. The relevant limit of finite trees is not the infinite homogenous tree graph $ \mathbb{T} $, but rather a single-ended canopy graph. On this tree graph, the random operator $ H_\lambda(\omega) $ is proven to have only pp spectrum at any $\lambda \geq 0 $. \\

In view of this negative result, which is caused by the presence of a large surface in the truncated tree, one needs to look at other finite graphs which are locally tree like. 
One example is the ensemble of {\bf random $(K+1)$-regular graphs}. It consists of the uniform probability measure on graphs on $ \mathbb{N} $ vertices where each vertex has $ (K+1) $ neighbors.\cite{Bol} As $ N\to \infty $ the girth of the graph, i.e. the minimal loop length, diverges in probability.  In fact, numerical simulations suggest that for large $ K $ the eigenvalue spacing distribution of the operator $ T $ approaches that of GOE.\cite{JM+99}

The ensemble of random regular graphs is hence well suited to investigate the spectral statistics conjecture of the operator $ H_\lambda(\omega) $ on the Bethe lattice. The gap statistics of the process of eigenvalues as well as the localization properties of the random eigenfunctions are currently under investigation in the physics community. There is numerical evidence\cite{BT} that the delocalized phase $ \{ (E,\lambda) \, | \, E \in \Sigma_\lambda^{ac} \} $ decomposes into two distinct regimes:
\begin{itemize}
\item One regime $(E, \lambda) $ (the so-called non-ergodic phase) in which the eigenvalue statistics is Poisson and the eigenfunctions occupy only a fraction of the graph. 
\item  another regime $(E, \lambda) $ (the so-called ergodic phase)  where the eigenvalue statistics is GOE and the eigenfunctions occupy the whole graph. 
\end{itemize}
A proof of even part of this exciting proposal remains a challenge.

\section*{Acknowledgments} 
This work is partially supported by the U.S. National Science Foundation under grant  DMS-0701181.

\bibliographystyle{ws-procs975x65}

\end{document}